\documentclass[superscriptaddress,groupedaddress,nofootnoteinbib,11pt,notoc]{article}
\pdfoutput=1
\usepackage{jcappub}

\usepackage{graphicx,color}
\usepackage{appendix}
\usepackage{latexsym,amsmath,amssymb,graphicx,booktabs}
\usepackage{epsfig,latexsym}
\usepackage{hyperref}
\numberwithin{equation}{section}% numera le equazioni seconde le sezioni , e.g. 1.15 invece che consecutivamente; anche le appendici, eq.~(A.1) etc. Richiede amsmath
\usepackage{url}

\definecolor{MyBlue}{rgb}{0.15,0.15,0.70}

\hypersetup{
colorlinks=true,
citecolor=MyBlue,
linkcolor=MyBlue,
urlcolor=MyBlue
}

\definecolor{lightgray}{gray}{0.9}

\newcommand{\dgw}{d_L^{\,\rm gw}}
\newcommand{\dem}{d_L^{\,\rm em}}

\newcommand{\iBox}{\Box^{-1}}

\renewcommand\({\left(}
\renewcommand\){\right)}
\renewcommand\[{\left[}
\renewcommand\]{\right]}

\newcommand{\ra}{\rightarrow}

\def\lsim{\raise 0.4ex\hbox{$<$}\kern -0.8em\lower 0.62
ex\hbox{$\sim$}}

\def\gsim{\raise 0.4ex\hbox{$>$}\kern -0.7em\lower 0.62
ex\hbox{$\sim$}}

\def\lbar{{\hbox{$\lambda$}\kern -0.7em\raise 0.6ex
\hbox{$-$}}}

\newcommand\eq[1]{eq.~(\ref{#1})}
\newcommand\eqs[2]{eqs.~(\ref{#1}) and (\ref{#2})}

\newcommand\pa{\partial}
\newcommand\p{\partial}

\newcommand\ee{\end{equation}}
\newcommand\be{\begin{equation}}
\def\bea{\begin{array}}
\def\eea{\end{array}}\def\ea{\end{array}}
\newcommand\ees{\end{eqnarray}}
\newcommand\bees{\begin{eqnarray}}

\def\dslash{\hspace{-1mm}\not{\hbox{\kern-2pt $\partial$}}}
\def\Dslash{\not{\hbox{\kern-2pt $D$}}}
\def\pslash{\not{\hbox{\kern-2.1pt $p$}}}
\def\kslash{\not{\hbox{\kern-2.3pt $k$}}}
\def\qslash{\not{\hbox{\kern-2.3pt $q$}}}

\def\p1{{\bf p}_1}
\def\p2{{\bf p}_2}
\def\k1{{\bf k}_1}
\def\k2{{\bf k}_2}

\newcommand{\gmn}{g_{\mu\nu}}

\newcommand{\Gmn}{G_{\mu\nu}}

\newcommand{\Tmn}{T_{\mu\nu}}
\newcommand{\Smn}{S_{\mu\nu}}

\newcommand{\dddM}{\kern 0.2em \raise 1.9ex\hbox{$...$}\kern -1.0em \hbox{$M$}}
\newcommand{\dddQ}{\kern 0.2em \raise 1.9ex\hbox{$...$}\kern -1.0em \hbox{$Q$}}
\newcommand{\dddI}{\kern 0.2em \raise 1.9ex\hbox{$...$}\kern -1.0em\hbox{$I$}}
\newcommand{\dddJ}{\kern 0.2em \raise 1.9ex\hbox{$...$}\kern-1.0em
\hbox{$J$}}
\newcommand{\dddcalJ}{\kern 0.2em \raise 1.9ex\hbox{$...$}\kern-1.0em
\hbox{${\cal J}$}}

\newcommand{\dddO}{\kern 0.2em \raise 1.9ex\hbox{$...$}\kern -1.0em
\hbox{${\cal O}$}}
\def\dddz{\raise 1.5ex\hbox{$...$}\kern -0.8em \hbox{$z$}}
\def\dddd{\raise 1.8ex\hbox{$...$}\kern -0.8em \hbox{$d$}}
\def\dddbd{\raise 1.8ex\hbox{$...$}\kern -0.8em \hbox{${\bf d}$}}
\def\ddbd{\raise 1.8ex\hbox{$..$}\kern -0.8em \hbox{${\bf d}$}}
\def\dddx{\raise 1.6ex\hbox{$...$}\kern -0.8em \hbox{$x$}}
\newcommand{\oma}{\Omega_{M}}
\newcommand{\ora}{\Omega_{R}}

\newcommand{\ola}{\Omega_{\Lambda}}

\newcommand{\rde}{\rho_{\rm DE}}
\newcommand{\wde}{w_{\rm DE}}

\graphicspath{ {./Graphs/} }

\title{Nonlocal gravity and gravitational-wave observations}
\author[a]{Enis Belgacem,}
\author[b]{Yves Dirian,}
\author[a]{Andreas Finke,}
\author[a]{Stefano Foffa,}
\author[a]{Michele Maggiore,}

\affiliation[a]{D\'epartement de Physique Th\'eorique and Center for Astroparticle Physics,\\
Universit\'e de Gen\`eve, 24 quai Ansermet, CH--1211 Gen\`eve 4, Switzerland}
\affiliation[b]{Center for Theoretical Astrophysics and Cosmology, Institute for Computational Science,
University of Z\"urich, CH-8057 Z\"urich, Switzerland}

\abstract{We discuss a modified gravity model which  fits cosmological observations at a level statistically indistinguishable from $\Lambda$CDM and  at the same time predicts very large deviations from General Relativity (GR) in the propagation of gravitational waves (GWs) across cosmological distances. The model is a variant of the RT nonlocal model proposed and developed by our group, with initial conditions set during inflation, and predicts a GW luminosity distance that, at the redshifts accessible to   LISA or to a third-generation GW detector such as the Einstein Telescope (ET),  can   differ from that in GR by as much as  $60\%$. An  effect of this size could be detected with just a single standard siren with counterpart by LISA or ET.  At the redshifts accessible to a LIGO/Virgo/Kagra network at target sensitivity the effect is smaller but still potentially detectable. Indeed, for the recently announced LIGO/Virgo NS-BH candidate S190814bv, the RT model predicts that, given the measured GW luminosity distance, the actual luminosity distance, and the redshift of an electromagnetic counterpart, would be smaller by as much as $7\%$ with respect to the  value inferred from $\Lambda$CDM.}

\begin{document}
\maketitle
\flushbottom

\section{Introduction}

The search for  deviations from  $\Lambda$CDM and of modifications of gravity at cosmological scales has been a main theme of
cosmological observations over the last  decades.  Modified gravity theories give rise to a different cosmological evolution for  the background,  usually encoded in the  dark energy (DE) equation of state $\wde(z)$,
and  to different  scalar and  tensor cosmological perturbations. Observations indicate that, for the background evolution and  scalar perturbations,  deviations cannot exceed a few percent (although a most notable discrepancy exists between  the local measurement of $H_0$ and the value inferred by  {\em Planck}  assuming $\Lambda$CDM~\cite{Riess:2019cxk}). 
For instance, at the background level, using the simple parametrization $\wde(z)=w_0$,  {\em Planck} 2015  combined with other datasets gives $w_0=-1.006\pm 0.045$~\cite{Planck_2015_CP}, i.e. the deviation from $\Lambda$CDM is bounded at the $4.5\%$ level.  Similarly, the DES Y1 results~\cite{Abbott:2018xao} put bounds at the level of $7\%$ on   the deviation  of  scalar perturbations from  $\Lambda$CDM. 
Tensor perturbations, i.e. GWs propagating on a cosmological background, are instead still a rather virgin territory, that we are beginning to explore thanks to the extraordinary LIGO/Virgo observations~\cite{Abbott:2016blz,TheLIGOScientific:2017qsa,LIGOScientific:2018mvr,Monitor:2017mdv}.
 
In GR the propagation of GWs is governed by  the equation
\be\label{4eqtensorsect}
\tilde{h}''_A+2{\cal H}\tilde{h}'_A+k^2\tilde{h}_A=0\, ,
\ee
where $A=\{+,\times\}$,  $h'=\pa_{\eta}h$,  $\eta$ is conformal time, 
${\cal H}=a'/a$ and  $a(\eta)$ is the  scale factor. 
In  modified gravity this equation is in general different.  However, 
a modification of  the coefficient of the $k^2$ term gives a speed of GWs different from that of light, which is now  excluded  at a level  $O(10^{-15})$ \cite{Monitor:2017mdv}, and indeed this has ruled  out a large class of  modifications of GR~\cite{Creminelli:2017sry,Sakstein:2017xjx,Ezquiaga:2017ekz,Baker:2017hug}.
Still, in many  modified gravity theories GW propagation is governed by the equation   
\be\label{prophmodgrav}
\tilde{h}''_A  +2 {\cal H}[1-\delta(\eta)] \tilde{h}'_A+k^2\tilde{h}_A=0\, ,
\ee
with some function $\delta(\eta)$~\cite{Saltas:2014dha,Gleyzes:2014rba,Lombriser:2015sxa,Nishizawa:2017nef,Arai:2017hxj,Belgacem:2017ihm,Amendola:2017ovw,Belgacem:2018lbp,Linder:2018jil,Lagos:2019kds,Nishizawa:2019rra,Belgacem:2019pkk}.  An important consequence of this modification is that, in the waveform of a coalescing binary, the  factor $1/d_L^{\,\rm em}(z)$, where $d_L^{\,\rm em}(z)$ is the standard luminosity distance  measured by electromagnetic probes, is  replaced by a `GW luminosity'  distance $d_L^{\,\rm gw}(z)$ (similar effects take place in theories with extra dimensions~\cite{Deffayet:2007kf,Pardo:2018ipy}).  The relation between $d_L^{\,\rm gw}(z)$ and $d_L^{\,\rm em}(z)$
is~\cite{Belgacem:2017ihm}
\be\label{dLgwdLem}
d_L^{\,\rm gw}(z)=d_L^{\,\rm em}(z)\exp\left\{-\int_0^z \,\frac{dz'}{1+z'}\,\delta(z')\right\}\, .
\ee
For  comparing with observations  it is useful to have a parametrization of the effect in terms of a small number of parameters, rather than a full function $\delta(z)$. A very convenient parametrization, in terms of two parameters $(\Xi_0,n)$,  has been proposed in \cite{Belgacem:2018lbp},
\be\label{eq:fit}
\frac{d_L^{\,\rm gw}(z)}{d_L^{\,\rm em}(z)}=\Xi_0 +\frac{1-\Xi_0}{(1+z)^n}\, .
\ee
This parametrization reproduces the fact that, as $z\ra 0$, $d_L^{\,\rm gw}/d_L^{\,\rm em}\ra 1$  since,  as the redshift of the source goes to zero, there can be no effect from modified propagation. 
In the limit of large redshifts, in \eq{eq:fit} $d_L^{\,\rm gw}/d_L^{\,\rm em}$ goes to a constant value $\Xi_0$. This is motivated by the fact that
in  typical  DE models the deviations from GR only appear in the recent cosmological epoch, so $\delta(z)$ goes to zero at large redshift and, from \eq{dLgwdLem},
$d_L^{\,\rm gw}(z)/d_L^{\,\rm em}(z)$  saturates to a constant. Indeed, in ref.~\cite{Belgacem:2019pkk} have been worked out the predictions of some of the best-studied modified gravity models such as several examples of
Horndeski and DHOST theories, non-local infrared modifications of gravity, or  bigravity theories. It has been found that  all these models (except bigravity, where there are non-trivial  oscillations due to the interaction between the two metrics) predict a propagation equation of the form (\ref{prophmodgrav}), with  functions $\delta(z)$ such that $d_L^{\,\rm gw}(z)/d_L^{\,\rm em}(z)$ is very well fitted by \eq{eq:fit}.

The parametrization (\ref{eq:fit}) has been used in \cite{Belgacem:2018lbp,Belgacem:2019tbw}
to forecast the sensitivity to modified GW propagation of  a network  of  detectors composed by   advanced LIGO/Virgo/Kagra  (HLVKI) at their target sensitivity, and of a third generation detectors such as the Einstein Telescope (ET), using as standard sirens the coalescence of binary neutron stars  (BNS) with an observed electromagnetic counterpart.
Using  state-of-the-art mock catalogs for the GW events and for the detection of an associated GRB, it is estimated that the HLVKI network can reach an accuracy on $\Xi_0$ of order $20\%$ over a few years of data taking, while ET could measure 
it to  about $1\%$.
In \cite{Belgacem:2019pkk} has been studied  the sensitivity of LISA, using as standard sirens  the coalescence of supermassive  black holes (SMBH), and it has been found that $\Xi_0$  could be measured  to 
$(1-4)\%$ accuracy, depending on assumptions on the population of 
SMBH black hole binaries. 

The crucial question is therefore what is the  size of the deviations  from the GR value $\Xi_0=1$, that one could expect from viable modified gravity models. As  mentioned above, for such models the deviations from $\Lambda$CDM  are bounded at the level of  $(4-5)\%$ for the background evolution and  about $7\%$  for  scalar perturbations, so one might expect that the  deviations in the tensor sector will be of the same order. However, below we will  present a model that is fully viable,  and which nevertheless can give deviations as large as $60\%$ in the tensor sector.
This is  excellent news for GW detectors, since it means that the new window on the Universe which is being opened by GW observations might reserve surprises from the point of view of cosmology, to the extent that advanced GW detectors could become the best instruments for detecting deviations from GR at cosmological scales.

\section{The RT nonlocal gravity model} 

Among the plethora of existing modified gravity models, a class that has been much developed in the last few years are nonlocal infrared modifications of gravity. The underlying idea, that in different forms goes back to  old works~\cite{Taylor:1989ua,Antoniadis:1986sb,Antoniadis:1991fa,Tsamis:1994ca,Polyakov:2007mm}, is that   quantum gravity  at large distances could induce cosmological effects, related to the emergence of  IR divergences in spacetimes of cosmological interest such as  de~Sitter space. These quantum effects generate nonlocal terms  in  the quantum effective action. This could  lead to the appearance of terms involving the inverse of the d'Alembertian operator, $\iBox$, that are relevant in the  infrared and  therefore affect the cosmological evolution. A first-principle understanding of these infrared effects is currently very difficult, so the work in this direction has been mostly of phenomenological nature, trying to identify models with interesting cosmological properties. However, already building a viable model turns out to be highly nontrivial.
A nonlocal gravity model based on this idea was proposed in \cite{Deser:2007jk} (see also 
\cite{Wetterich:1997bz} for earlier work), but is now ruled out~\cite{Belgacem:2018wtb}. A more recent  twist of the idea is that infrared effects might generate dynamically a mass, associated to nonlocal terms. The first viable model of this type was proposed 
in \cite{Maggiore:2013mea}, elaborating on earlier work in \cite{ArkaniHamed:2002fu,Jaccard:2013gla}, and is defined  by  the nonlocal equation of motion
\be\label{RT}
\Gmn -(m^2/3)\(\gmn\iBox R\)^{\rm T}=8\pi G\,\Tmn\, .
\ee
Here $m$ is a mass scale, eventually taken of order $H_0$, that replaces the cosmological constant, and 
the superscript `T' denotes the extraction of the transverse part of a tensor, based on the fact that any   symmetric tensor $\Smn$ can  be decomposed as 
$S_{\mu\nu}=S_{\mu\nu}^{\rm T}+(1/2)(\nabla_{\mu}S_{\nu}+\nabla_{\nu}S_{\mu})$,
where the transverse part $S_{\mu\nu}^{\rm T}$  satisfies
$\nabla^{\mu}S_{\mu\nu}^{\rm T}=0$.  We refer to it as the `RT' model, where R stands for the  Ricci scalar and T for the extraction of the transverse part of $S_{\mu\nu}\equiv \gmn\iBox R$.
Detailed discussions of the reasoning that led to this specific structure (that corresponds to a dynamical mass generation for the conformal mode), of  conceptual aspects related to the appearance of these nonlocal terms (that respect causality and do not introduce  extra degrees of freedom), and of its cosmological consequences can be found in the reviews~\cite{Maggiore:2016gpx,Belgacem:2017cqo}. Here we limit ourselves to recalling that several studies  \cite{Maggiore:2013mea,Foffa:2013vma,Kehagias:2014sda,Maggiore:2014sia,Nesseris:2014mea,Dirian:2014ara,Barreira:2014kra,Dirian:2014bma,Dirian:2016puz,Dirian:2017pwp,Belgacem:2017cqo,Belgacem:2018wtb} have shown that the RT model has a viable cosmological background evolution, where the nonlocal term acts as an effective dark energy density and drives  accelerated expansion in the recent epoch; it has stable cosmological perturbations in the scalar and  tensor sectors (a nontrivial condition that ruled out many modified gravity models); tensor perturbations propagate at the speed of light;   the model fits CMB, BAO, SNe and structure formation  at a level statistically equivalent to $\Lambda$CDM; and it reproduces the successes of GR at solar system and laboratory scales.  Studies of  possible variations of the idea have singled out the RT model as the only known nonlocal model that passes all these  tests.

\begin{figure}[t]
 \centering
 \includegraphics[width=0.46\textwidth]{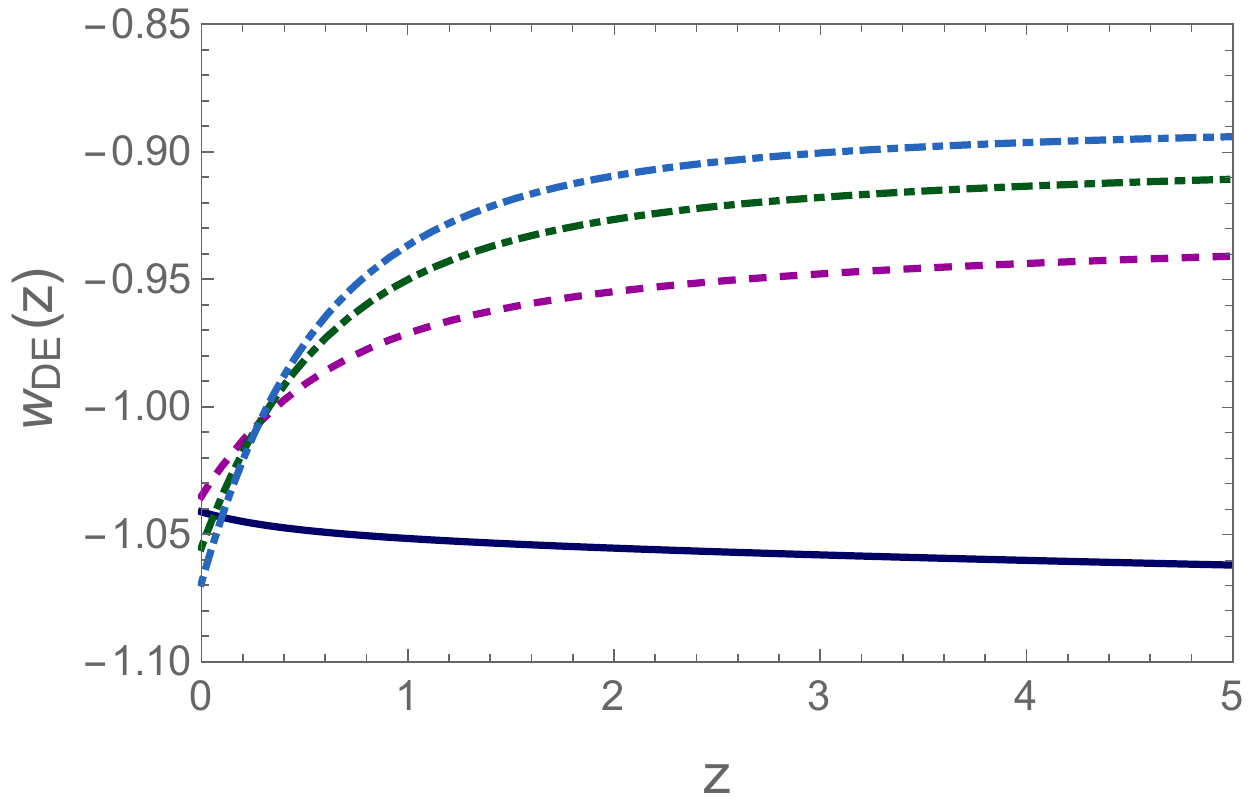}
 \caption{The DE equation of state $\wde(z)$ as a function of redshift for the minimal RT model (blue solid line), for $\Delta N=34$ (magenta, dashed), $\Delta N=50$ (green, dot-dashed) and $\Delta N=64$ (cyan, dot-dashed).}
  \label{fig:wde}
\end{figure}

A point that will be important in the following is that nonlocal models have a hidden freedom related to the choice of initial conditions for some auxiliary fields that are introduced to write the equations of motion in local form (or, equivalently, there is an implicit freedom in the definition of  $\iBox$). Once again, we refer the reader to ~\cite{Maggiore:2013mea,Maggiore:2016gpx,Belgacem:2017cqo} for full discussions. However, the bottom line is that, if one had an explicit derivation of the effective nonlocal terms  of the quantum effective action from the fundamental underlying local theory, one could derive these initial conditions explicitly in terms of those of the metric. In the absence of such a  derivation, these initial conditions must be treated as phenomenological parameters. This, however, does not introduce as much freedom as one might fear. Indeed,
if we set initial conditions ${\cal O}(1)$ during  radiation dominance (RD), the evolution is fully determined because the freedom in the initial conditions  either corresponds to re-introducing a cosmological constant, that we set to zero because, at least in the minimal setting, we want to reproduce the accelerated expansion without reintroducing a cosmological constant, or else corresponds to irrelevant directions in parameter space (the same happens at the level of cosmological perturbations). The situation is different if we start the evolution in a primordial inflationary phase; in this case, because of the presence of a growing mode,  if we start with initial conditions of order one at some  time  when there are still $\Delta N$ e-folds until the end of  inflation, an auxiliary field gets a value of order $e^{\alpha \Delta N}$ when inflation ends and RD begins, with $\alpha$  a known constant. This instability has no effect on the evolution during inflation (neither at the background level not for the perturbations), since the energy scale associated to the nonlocal term is negligible with respect to the inflationary scale. However,
the subsequent evolution during RD inherits a dependence on $\Delta N$. Full details are given in \cite{Maggiore:2016gpx,Cusin:2016mrr}.
The model that we  study below is indeed the RT model with initial conditions  set during inflation. For inflation taking place at a scale $M$, assuming instantaneous reheating,  the minimum number of efolds required to solve the flatness and horizon problems is 
\be
\Delta N\simeq 64-\log\frac{10^{16}\, {\rm GeV}}{M}\, .
\ee
In the following, beside the `minimal' model defined by initial conditions of order one during RD, which corresponds to $\Delta N=0$, we will study also the cases 
$\Delta N=\{34,50,64\}$ that, in the above approximation, correspond to  $M=\{10^3,10^{10},10^{16}\}$~GeV, respectively.
The  result for the background evolution, i.e. for $\wde(z)$, were already given  in 
\cite{Maggiore:2016gpx}, and we reproduce them in Fig.~\ref{fig:wde}. In the minimal model $\wde$ is always on the phantom side, $\wde(z)<-1$, while for large $\Delta N$ it evolves from a non-phantom value at large $z$ toward a phantom value today, with phantom crossing near $z\simeq 0.30-0.35$. In any case at $z<1$, when DE becomes important, all these curves are  within about $5\%$  of the $\Lambda$CDM value $-1$, so  the background evolution of these models is still quite  close to that of $\Lambda$CDM. 

\begin{table*}[t]
\centering
\begin{tabular}{|l||c|c|c|c|c|}
\hline
%\multicolumn{1}{|l||}{ } & \multicolumn{5}{|c|}{CMB+BAO+SNe} \\ \hline
Parameter & $\nu\Lambda$CDM & RT, minimal & RT, $\Delta N=34$ & RT, $\Delta N=50$ & RT, $\Delta N=64$ \\ \hline
$H_0$ \phantom{\Big|}& $67.60^{+0.66}_{-0.55}$ & $68.35^{+0.75}_{-0.71}$ & $67.68^{+0.67}_{-0.59}$ & $67.71^{+0.56}_{-0.62}$ & $67.66^{+0.68}_{-0.64}$\\
$\sum_{\nu}m_{\nu}\ [{\rm eV}]$ & $< 0.10$ (at $1\sigma$) & $0.126^{+0.055}_{-0.101}$ & $< 0.10$ (at $1\sigma$) & $< 0.08$ (at $1\sigma$) &$< 0.09$ (at $1\sigma$)\\
$\omega_c$ \phantom{\Big|} & $0.1189^{+0.0011}_{-0.0011}$ & $0.1194^{+0.0012}_{-0.0012}$ & $0.1186^{+0.0012}_{-0.0012}$ & $0.1185^{+0.0012}_{-0.0011}$ & $0.1184^{+0.0012}_{-0.0012}$\\
100$\omega_b$ \phantom{\Big|} & $2.229^{+0.014}_{-0.015}$ & $2.225^{+0.015}_{-0.015}$ & $2.230^{+0.016}_{-0.014}$ & $2.231^{+0.015}_{-0.016}$ & $2.232^{+0.016}_{-0.016}$\\
$\ln (10^{10} A_s)$\phantom{\Big|} & $3.071^{+0.026}_{-0.029}$ & $3.070^{+0.029}_{-0.032}$ & $3.076^{+0.027}_{-0.031}$ & $3.075^{+0.027}_{-0.028}$ & $3.080^{+0.028}_{-0.029}$\\
$n_s$ \phantom{\Big|} & $0.9661^{+0.0043}_{-0.0043}$ & $0.9648^{+0.0045}_{-0.0043}$ & $0.9670^{+0.0045}_{-0.0045}$ & $0.9670^{+0.0042}_{-0.0046}$ & $0.9673^{+0.0046}_{-0.0048}$\\
$\tau_{\rm re}$ \phantom{\Big|} & $0.06965^{+0.01393}_{-0.01549}$ & $0.06858^{+0.01534}_{-0.01721}$ & $0.07257^{+0.01491}_{-0.01585}$ & $0.07183^{+0.01430}_{-0.01518}$ & $0.07462^{+0.01488}_{-0.01609}$\\
\hline
$\Omega_M$\phantom{\Big|} & $0.3109_{-0.0084}^{+0.0069}$ & $0.3061_{-0.0091}^{+0.0079}$ & $0.3095_{-0.0081}^{+0.0077}$ & $0.3087_{-0.0074}^{+0.0075}$ & $0.3091_{-0.0086}^{+0.0077}$\\
$z_{\rm re}$ & $9.150_{-1.355}^{+1.396}$ & $9.058_{-1.487}^{+1.587}$ & $9.417_{-1.376}^{+1.429}$ & $9.349_{-1.279}^{+1.402}$ & $9.604_{-1.467}^{+1.402}$\\
$\sigma_8$ \phantom{\Big|} & $0.8157^{+0.0135}_{-0.0104}$ & $0.8196^{+0.0165}_{-0.0130}$ & $0.8162^{+0.0140}_{-0.0112}$ & $0.8164^{+0.0128}_{-0.0112}$ & $0.8166^{+0.0129}_{-0.0114}$\\
\hline
%$\chi^2$ \phantom{\Big|} & 13630.78 & 13631.54 & 13629.68 & 13629.58 & 13629.60 \\
$\Delta\chi^2$\phantom{\big|} & 0 & 0.76 & -1.10 & -1.20 & -1.18\\
\hline
\end{tabular}
\caption{\label{tab:results} Mean values (with $1\sigma$ errors) of the parameters  for 
$\nu\Lambda$CDM  and the RT model with $\Delta N$= 0, 34, 50, 64, using CMB, BAO and SNe. $H_0$ is in units of
${\rm km}\, {\rm s}^{-1}\, {\rm Mpc}^{-1}$. The last line gives the difference in the $\chi^2$ of each given model with respect to $\nu\Lambda$CDM. The RT model with $\Delta N=50$ or with $\Delta N=64$ fit the data slightly better than $\nu\Lambda$CDM, but the difference is not statistically significant.}
\end{table*}

We then study whether this evolution, together with the associated scalar perturbations, is consistent with observations.
Cosmological perturbations for the RT model were computed in~\cite{Dirian:2014ara} and, for the minimal model, were implemented into a Markov Chain Monte Carlo (MCMC) and compared with data in
\cite{Dirian:2014bma,Dirian:2016puz}. We extend here  the  analysis to $\Delta N=34,50,  64$. We use the same  datasets as in our previous works, namely the 2015 {\em Planck} data, JLA supernovae and a compilation of BAO, and we vary the standard set of cosmological parameters (see \cite{Belgacem:2017cqo} for details), including the sum of neutrino masses. The results are shown in Table~\ref{tab:results}. From the mean values of the parameters we see that the RT model is very close to 
$\nu\Lambda$CDM (i.e. $\Lambda$CDM where  the neutrino masses are allowed to vary), particularly for large $\Delta N$. In the last line we give   the difference of the $\chi^2$ of a given model, with respect to  $\nu\Lambda$CDM. According to  the conventional Jeffreys' scale, a difference $|\Delta \chi^2| \leq  2$ implies  statistically equivalence between the two models compared, while $|\Delta \chi^2| \gtrsim 2$ suggests ``weak evidence'', and $|\Delta \chi^2|\gtrsim 6$ indicates ``strong evidence''.  
Thus, all models shown in Table~\ref{tab:results} fit the data at a statistically equivalent level. Observe also that the RT model does not relieve the tension with the local $H_0$ measurement, particularly at large $\Delta N$. Indeed, recent work  indicates that  it might not be possible to solve  the $H_0$ tension solely with a modification
of the late-Universe dynamics~\cite{Poulin:2018zxs,Aylor:2018drw}.

\begin{figure}[t]
 \centering
 \includegraphics[width=0.46\textwidth]{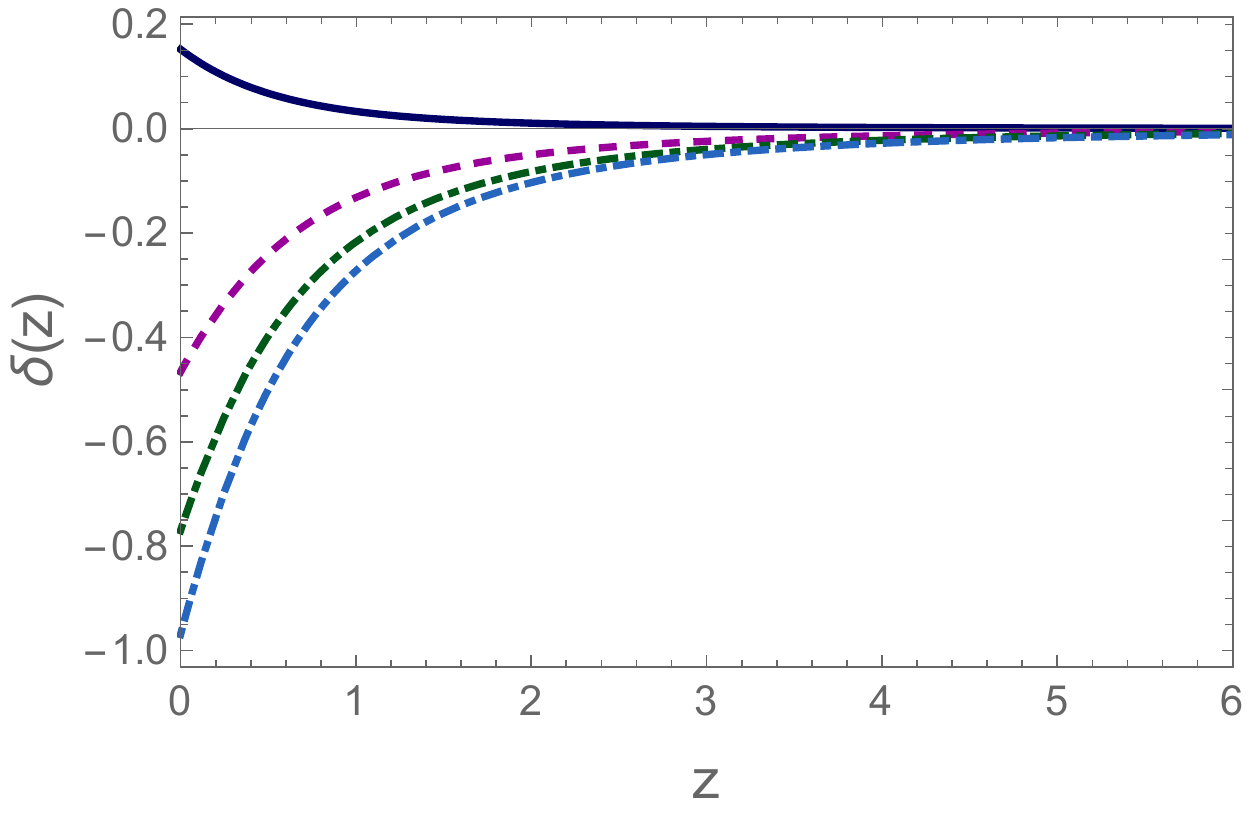}
 \qquad
  \includegraphics[width=0.46\textwidth]{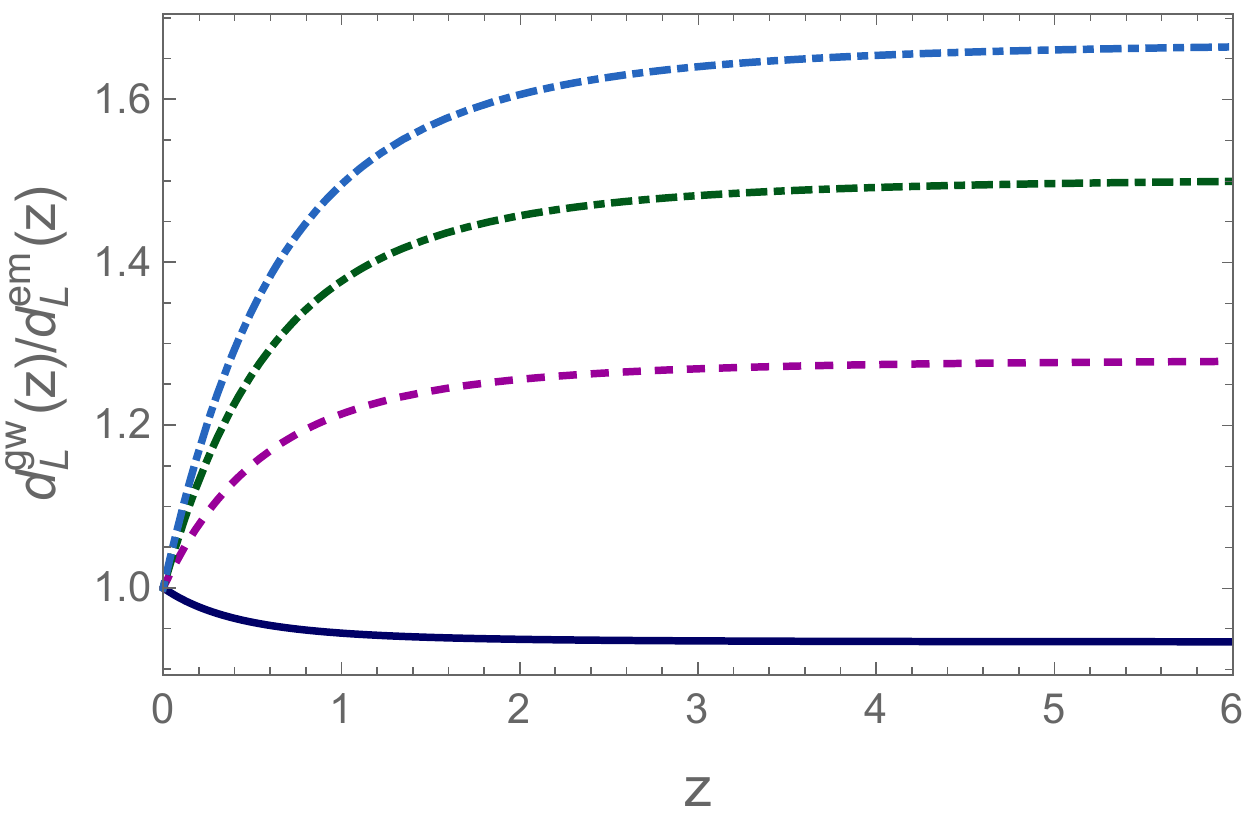}
 \caption{The functions $\delta(z)$ (left panel) 
 and  $d_L^{\,\rm gw}(z)/d_L^{\,\rm em}(z)$ (right panel)  for the minimal RT model (blue solid line), for $\Delta N=34$ (magenta, dashed), $\Delta N=50$ (green, dot-dashed) and $\Delta N=64$ (cyan, dot-dashed).}
  \label{fig:delta_vs_z}
\end{figure}

%\begin{figure}[t]
% \centering
% \includegraphics[width=0.46\textwidth]{dgw_su_dem.pdf}
% \caption{As in Fig.~\ref{fig:delta_vs_z}, for $d_L^{\,\rm gw}(z)/d_L^{\,\rm em}(z)$.}
%  \label{fig:dgw_su_dem}
%\end{figure}

We next discuss  tensor perturbations. The equation for tensor perturbations in the RT model is of the form (\ref{prophmodgrav}), with $\delta(\eta)$ determined by the evolution of the auxiliary fields of the model~\cite{Dirian:2016puz}. Performing the numerical integration of the equations of motion we obtain $\delta(z)$, shown in the left panel of Fig.~\ref{fig:delta_vs_z}.
Using \eq{dLgwdLem}
we then obtain  the ratio $d_L^{\,\rm gw}(z)/d_L^{\,\rm em}(z)$, which is shown in  the right panel of 
 Fig.~\ref{fig:delta_vs_z}  (the plot for the minimal model was already shown in \cite{Belgacem:2019pkk}). The result is quite surprising: for large $\Delta N$, $d_L^{\,\rm gw}(z)/d_L^{\,\rm em}(z)$ becomes larger than one, corresponding to a smaller GW amplitude compared to the GR prediction. What is most remarkable, however, is the size of the deviation from the GR value $\Xi_0=1$, that in the minimal model was about $6.6\%$, while for  $\Delta N=64$ is now at the level of $65\%$, one order of magnitude larger! 
 All the curves are perfectly fitted by \eq{eq:fit}. Keeping $\oma$ and $H_0$ fixed at the values $\oma=0.30$ and $H_0=68.80$ and varying only $\Delta N$ we get the values shown in Table~\ref{tab:Xi0nfixed}.
If instead for each model we use its own mean values  of the parameters from Table~\ref{tab:results}, we find the results shown in Table~\ref{tab:Xi0nbest}.

\begin{table*}[t]
\centering
\begin{tabular}{|l||c|c|c|c|}
\hline
Parameter & RT, minimal & RT, $\Delta N=34$ & RT, $\Delta N=50$ & RT, $\Delta N=64$ \\ \hline
$\Xi_0$      &0.93 & 1.28  &1.50  &1.67\\
$n$             &2.59 & 2.07  &1.99  &1.94\\
$\delta(0)$ &0.15  &-0.47 &-0.77 &-0.97\\
\hline
\end{tabular}
\caption{\label{tab:Xi0nfixed} The predictions for $
\Xi_0$, $n$  and $\delta(z=0)$ of the RT model  with $\Delta N= 0, 34, 50, 64$ 
keeping $\oma$ and $H_0$ fixed at the values $\oma=0.30$ and $H_0=68.80$ and varying only $\Delta N$.}
\end{table*}

\begin{table*}[t]
\centering
\begin{tabular}{|l||c|c|c|c|}
\hline
Parameter & RT, minimal & RT, $\Delta N=34$ & RT, $\Delta N=50$ & RT, $\Delta N=64$ \\ \hline
$\Xi_0$      &0.93             & 1.27                        &1.49                         &1.65\\
$n$             &2.59             & 2.08                        &2.00                         &1.95\\
$\delta(0)$ &0.15             &-0.46                       &-0.76                         &-0.95\\
\hline
\end{tabular}
\caption{\label{tab:Xi0nbest} As in Table~\ref{tab:Xi0nfixed}, using for each $\Delta N$ the mean values for $\oma$ and $H_0$ given in Table~\ref{tab:results}.}
\end{table*}
\section{Detectability of the effect} 

\subsection{Sensitivity to $\Xi_0$ of 2G and 3G detectors}

The estimates of detector sensitivities to $\Xi_0$ mentioned above show that deviations from GR at this level are measurable at 3G detectors, and possibly even at 2G detectors.  
The analysis of ref.~\cite{Belgacem:2019tbw}, using
mock catalogs of joint GW-GRB  detections and assuming the functional form (\ref{eq:fit}), which fits perfectly the prediction of the RT model, shows that  the HLVKI network in 10~yr of run could measure $\Xi_0$ to an accuracy 
$\Delta \Xi_0\simeq 0.1$, while a shorter and more realistic time span, say 2-3 years, could still be enough to reach 
$\Delta \Xi_0\simeq 0.2$. This would be sufficient to detect with good confidence the difference between the prediction $\Xi_0=1.6$ of the $\Delta N=64$ scenario, and the GR value $\Xi_0=1$. Observe that in \cite{Belgacem:2019tbw} were only studied the coincidences of GW events with GRBs. For well-localized events,  several more counterparts could be obtained  with optical/IR telescopes.
Note  that, at the redshifts accessible to the HLVKI network, we are not yet in the asymptotic regime 
$d_L^{\,\rm gw}(z)/d_L^{\,\rm em}(z)\simeq \Xi_0$, a fact that is anyhow taken into account by the use of the parametrization (\ref{eq:fit}) in the analysis. Observe that  being able to detect events at  not-too-small redshifts is crucial here. 
In fact,  to first order in $z$, \eq{dLgwdLem} becomes
$d_L^{\,\rm gw}(z)/d_L^{\,\rm em}(z)\simeq 1-z \delta(0)$, so  the deviation from 1 is  linear in $z$. 
An event such as GW170817, 
at $z\simeq 0.01$, therefore gives a measure of $\delta(0)$ with an error   10 times larger than an event at $z=0.1$ with a comparable observational error. Since the overall error scales with  the number of similar events  roughly as $1/\sqrt{N}$,  a single event at $z=0.1$  contributes as much as ${\cal O}(100)$ events at $z=0.01$.

For a 3G detector such as ET the perspectives look extremely interesting.  ET could see hundreds
of BNS with  counterpart   to $z\simeq 2$, where $d_L^{\,\rm gw}(z)/d_L^{\,\rm em}(z)\simeq \Xi_0$, with an  error $\Delta d_L^{\,\rm gw}/d_L^{\,\rm gw}$  of order $(5-10)\%$ (see sect.~2.3.1 and Fig.~5 of \cite{Belgacem:2019tbw}). 
With this sensitivity,  already a single event with electromagnetic counterpart, or just a few of them, could be enough to discriminate between $\Xi_0\simeq 1.6$ and  $\Xi_0=1$. With the few hundreds of  joint GW-GRB detection estimated in \cite{Belgacem:2019tbw}, one can reach $\Delta\Xi_0/\Xi_0\simeq 1\%$ or better.

Also for  LISA the perspectives are quite exciting. From Table~2 of \cite{Belgacem:2019pkk}, LISA could measure $\Xi_0$ with an error
$\Delta\Xi_0\simeq\{0.023,0.036,0.044\}$ in three different scenarios for SMBH formations that lead to catalogs containing, respectively, $N=\{32,12,9\}$ events. These numbers are well reproduced by   $\Delta\Xi_0\simeq 0.13/\sqrt{N}$, so  each SMBH event gives a measure of $\Xi_0$ with an average accuracy of about $13\%$ (which becomes $6\%$ if we  use  a more optimistic scenario for the error on the redshift determination of the source). Thus, a single SMBH event at LISA could be sufficient to detect the effect.

It is remarkable that such a measurement would reveal the nature of dark energy, and at the same time would give information on primordial inflation through its dependence on $\Delta N$.

\subsection{Application to  the candidate NS-BH event  S190814bv}

It is very interesting to see how the above predictions apply to  the very recent candidate event S190814bv, that, at the time of writing, has just been  posted\footnote{The event has been posted on Aug. 14, 2019, about six weeks after the v1 version of this paper appeared on the arxiv.} on the public database of GW candidate events of the LIGO/Virgo collaborations~(see \url{https://gracedb.ligo.org/superevents/S190814bv/}). This event has  an extremely low false alarm rate (less than one false alarm per $10^{25}$~yr), and is currently classified as a 
neutron-star -- black-hole (NS-BH) coalescence with $>99\%$ probability.
At the time of writing, the information on the event has  just been disseminated to the network of electromagnetic telescopes, and the search for an electromagnetic counterpart is under way. Even 
if information on this candidate event is still preliminary, still
it is interesting to estimate the effect of modified GW propagation on an event of this type. Currently, the best estimate for the luminosity distance  of the event, based on the GW signal, is $d_L=(267\pm 52)\,  {\rm Mpc}$. In $\Lambda$CDM, this would be the same as the luminosity distance to the source measured by electromagnetic signals, while we have seen that in the RT nonlocal model (and more general, in basically all modified gravity models) there are two different notions of luminosity distance, the GW luminosity distance $\dgw$, which is the quantity measured by GW detectors, and the `electromagnetic' luminosity distance $\dem$ measured by electromagnetic observations. In $\Lambda$CDM, where the two notions coincide, we write simply $\dgw(z)=\dem(z)\equiv d_L(z)$. From a measurement of $d_L(z)$, the redshift of the source can be obtained by inverting the expression
\be\label{dLem}
d_L(z)=\frac{1+z}{H_0}\int_0^z\, 
\frac{d\tilde{z}}{\sqrt{\ora (1+\tilde{z})^4+\oma (1+\tilde{z})^3+\ola }}\, ,
\ee
where  $\ora$, $\oma$ and $\ola=1-\ora-\oma$  are the density fractions associated to radiation, matter and to the cosmological constant, respectively.  In contrast, in the RT model  the electromagnetic luminosity distance is
\be\label{dem}
\dem(z)=\frac{1+z}{H_0}\int_0^z\, 
\frac{d\tilde{z}}{\sqrt{\ora (1+\tilde{z})^4+\oma (1+\tilde{z})^3+\rde(\tilde{z})/\rho_0 }}\, ,
\ee
where $\rde(z)$ is the dark energy density in the RT model and  $\rho_0=3H_0^2/(8\pi G)$, and 
the 
GW luminosity distance is very well reproduced by
\be\label{dgw}
\dgw(z)= \[ \Xi_0 +\frac{1-\Xi_0}{(1+z)^n}\] \, \dem(z)\, .
\ee

The value of $z$ inferred from a measurement of $d_L(z)$ depends in particular on the value chosen for $H_0$. Given the well-known discrepancy, mentioned above, between the value obtained from  the local measurement of $H_0$ and that  from  {\em Planck} (combined with BAO and SNe), this gives a systematic error that is larger than the statistical error of each measurement, so the two cases must be treated separately. We first assume the correctness of the value of $H_0$ obtained from {\em Planck}+BAO+SNe. Then for $\Lambda$CDM 
we use the values $H_0=67.60^{+0.66}_{-0.55}$ (in units of
${\rm km}\, {\rm s}^{-1}\, {\rm Mpc}^{-1}$) and $\oma=0.3109$  from the first column of Table~\ref{tab:results}.\footnote{Note that, since we will get $z\simeq 0.06$, we are in the regime where $d_L\simeq H_0^{-1} z [1+{\cal O}(z)]$, so the precise value of $\oma$ has limited impact, and the error on $\oma$ has no influence on the error on $z$.} In this case we find that 
$d_L=(267\pm 52)\,  {\rm Mpc}$ corresponds to a source redshift
\be\label{z58}
z=  0.058\pm 0.011\, , \qquad (\mbox{$\Lambda$CDM,\,\, $H_0$ from {\em Planck}+BAO+SNe})\, ,
\ee
where the error quoted is the one induced by the error on the measurement of $d_L$. The statistical error on the measurement of $H_0$ induces a further error $\Delta z=z (\Delta H_0)/H_0\simeq 0.0005$, which is beyond the number of digits given in \eq{z58}.
Let us compare this result with that obtained in the RT model, using $\Delta N=64$ to maximize the effect of modified GW propagation. We then use the mean values for $H_0$ and $\oma$ obtained for the  RT model  (with $\Delta N=64$) from the 
{\em Planck}+BAO+SNe observations, given in the last column of Table~\ref{tab:results}. 
Using \eqs{dem}{dgw} and
inverting
$\dgw(z)=(267\pm 52)\,  {\rm Mpc}$,
we find that the redshift  of the source is
\be
z=0.054\pm 0.010\, , \qquad (\mbox{RT with $\Delta N=64$,\,\, $H_0$ from {\em Planck}+BAO+SNe})\, ,
\ee
and the actual (`electromagnetic')  luminosity distance of the source is
$\dem =(251 \pm 49)\, {\rm Mpc}$.

If instead we use the value $H_0=74.03\pm 1.42$ from local measurements~\cite{Riess:2019cxk},  in $\Lambda$CDM we get
\be\label{zRiessLCDM}
z=0.063\pm 0.012\, , \qquad (\mbox{$\Lambda$CDM},\,\, H_0=74.03)\, ,
\ee
while  the RT model predicts that the redshift of the source is rather
\be\label{zRiessRT}
z=0.059\pm 0.010\, , \qquad (\mbox{RT with $\Delta N=64$,\,\, $H_0=74.03$})\, ,
\ee
and its electromagnetic luminosity distance is $\dem=(250\pm 49)$~Mpc.
To both \eqs{zRiessLCDM}{zRiessRT} one should add  $\Delta z=z (\Delta H_0)/H_0\simeq 0.001$ from the error on $H_0$.
In both cases ($H_0$ from {\em Planck}+BAO+SNe, or $H_0$ from local measurements) the RT model with $\Delta N=64$ predicts that the actual cosmological redshift of the source  is smaller than the value that would be inferred using $\Lambda$CDM, by about $7\%$. This prediction can in principle be tested if an electromagnetic counterpart will be discovered. Of course, to test this prediction one must first be able to  arbitrate the tension between {\em Planck} and local measurements of $H_0$, which can be done with standard sirens themselves. However, it is quite remarkable that, in the RT model, even for a single event at the distance of  S190814bv,  the effect from modified GW propagation on the redshift, or equivalently on the luminosity distance, can be as large as $7\%$ (with this value reached for $\Delta N=64$, and smaller effects for smaller $\Delta N$), which is well above the statistical errors of both the {\em Planck} and the local measurements of $H_0$.

\vspace{5mm}\noindent
{\bf Acknowledgments.} 
The work  of E.B., A.F., S.F. and M.M. is supported by the  Swiss National Science Foundation and  by the SwissMap National Center for Competence in Research. The work of Y.D. is supported by  Swiss National Science Foundation and by  a Consolidator Grant of the European Research Council (ERC-2015-CoG grant 680886).

\clearpage

\bibliographystyle{utphys}
\bibliography{myrefs}

\end{document}